\documentclass[a4paper,11pt]{article}

\usepackage{jheppub} 

\usepackage{epstopdf} 
\usepackage[T1]{fontenc} 
\usepackage[dvipsnames]{xcolor}

\usepackage{graphics}
\usepackage[utf8]{inputenc}
\usepackage{xspace}
\usepackage{amsmath}
\usepackage{amssymb}
\usepackage{url}
\usepackage{rotating}
\usepackage{enumerate}
\usepackage{graphicx}
\usepackage{pdflscape, afterpage, capt-of} 
\usepackage{slashed} 
\usepackage[normalem]{ulem}
\usepackage{nicefrac}
\usepackage[export]{adjustbox} 
\usepackage{makecell} 
\usepackage{subcaption}

\usepackage{def} 

\newcommand{\private}[1]{} 

\title{\boldmath Orthogonal color bases for exotic representations}

\author[a]{Malin Sjodahl}

\usepackage{ytableau}
\ytableausetup{mathmode, centertableaux, boxsize=1.5ex}
\affiliation[a]{Department of Physics, Lund
  University, Box 118 221 00 Lund, Sweden}

\emailAdd{malin.sjodahl@fysik.lu.se}

\abstract{
  A complication in the treatment of any strongly charged particle is
  the SU(3) color structure. For the standard model quarks antiquarks and gluons
  there are various well-known strategies for dealing with the color structure,
  including orthogonal multiplet bases.
  For exotic representations, beyond triplets, antitriplets and octets,
  non-orthogonal bases have been systematically worked out only recently. In this letter
  we demonstrate how to construct orthogonal multiplet bases also
  for processes including exotic representations.
}

\begin{document} 
\preprint{MCNET-24-19}
\maketitle
\flushbottom

\section{Introduction}
\label{introduction}

The calculation of scattering amplitudes involving strongly charged
particles necessarily entails dealing with the SU(3) color structure.
For the representations of standard model particles, triplets, antitriplets,
octets and singlets, several strategies have been worked out over the
decades. Most naively one can of course simply sum over the color
indices appearing in SU(3) generators and structure
constants. For processes with many partons it is, however,
often advantageous to employ a color basis or a spanning set for the
color decomposition.
For the standard model representations, a common option
is trace bases, where gluons are attached to closed quark lines,
giving an actual trace, or to traces that have been cut open and
carry free quark indices \cite{Paton:1969je, Dittner:1972hm, Cvitanovic:1976am,Berends:1987cv, Mangano:1987xk, Mangano:1988kk, Kosower:1988kh, Nagy:2007ty, Sjodahl:2009wx, Alwall:2011uj, Sjodahl:2012nk,
  Sjodahl:2014opa, Platzer:2012np,Platzer:2018pmd,Frederix:2021wdv}.
It is also possible to translate all gluon
indices to (a linear combination) of quark lines, giving a color-flow basis
\cite{tHooft:1973alw,Kanaki:2000ms,Maltoni:2002mq, Kilian:2012pz, Platzer:2013fha,AngelesMartinez:2018cfz,DeAngelis:2020rvq,Platzer:2020lbr,Frederix:2021wdv}, where all color structure is expressed in terms
of only quark lines. While these ``bases'', strictly speaking,
rather are spanning sets, as the ``basis vectors'' form an overcomplete
set, it is of course perfectly possible to form orthogonal linear
combinations giving an orthogonal basis. Nothing stops a naive
Gram-Schmidt approach, but the resulting bases soon
become untractable, and over the years, instead, orthogonal multiplet bases
have been constructed, first for a few partons
\cite{MacFarlane:1968vc,Butera:1979na,Kyrieleis:2005dt,
  Dokshitzer:2005ig,Sjodahl:2008fz,Beneke:2009rj},
and later for an in principle arbitrary set of standard model partons
 \cite{Keppeler:2012ih,
  Du:2015apa,Sjodahl:2015qoa,Keppeler:2013yla,Alcock-Zeilinger:2016bss,
  Alcock-Zeilinger:2016sxc,Alcock-Zeilinger:2016cva,Sjodahl:2018cca}.

During the last years, there has been some development towards
constructing bases also for exotic representations.
Attempts for the sextet representation have been presented in Ref. \cite{PhysRevD.105.035014},
and a general procedure for (non-orthogonal) basis construction
was worked out and implemented in Ref. \cite{Ohl:2024fpq}.

In this brief letter, we first review the construction of orthogonal
multiplet bases in \secref{sec:multiplet bases}.
We then see how orthogonal multiplet bases involving exotic representations
can be constructed, by building on a type of multiplet bases where quarks,
antiquarks and gluons are attached to a backbone chain of
general representations. Finally we conclude, and make an outlook in \secref{sec:conclusion}.

\section{Multiplet bases}
\label{sec:multiplet bases}

A systematic approach to orthogonal bases is to
employ representation theory, and group particles together according to
which representation the set of particles transforms under.
The simplest non-trivial example is $q_1 q_2\rightarrow \q_3 q_4$,
giving the representations
\begin{eqnarray}
  \underbrace{\begin{ytableau}  \ \end{ytableau}}_{3}
  \otimes
  \underbrace{\begin{ytableau}  \ \end{ytableau}}_{3}
      &=&
      \underbrace{\begin{ytableau}  \ & \ \end{ytableau}}_{6}
        \oplus
        \underbrace{\begin{ytableau}  \  \\ \ \end{ytableau}}_{\overline{3}}\;,
\end{eqnarray}
where projectors onto overall sextet and antitriplet states are given
as indicated by the symmetries of the corresponding Young diagram,
\begin{eqnarray}
  \label{eq:projectors}
  \mathbf{P}^6
  &=&\frac{1}{2}\left(
  \raisebox{-0.5 \height}{\includegraphics[scale=0.35]{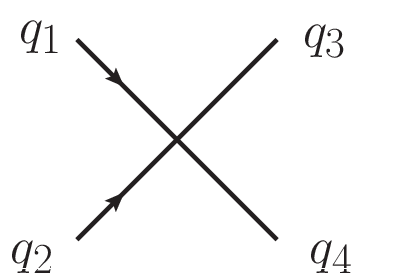}} +
  \raisebox{-0.5 \height}{\includegraphics[scale=0.35]{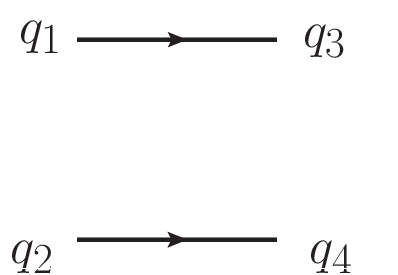}}
  \right)\nonumber \\
  \mathbf{P}^{\overline{3}}
  &=&\frac{1}{2}\left(
  \raisebox{-0.5 \height}{\includegraphics[scale=0.35]{jaxodraw/Crossed.eps}} -
  \raisebox{-0.5 \height}{\includegraphics[scale=0.35]{jaxodraw/Straight.eps}}
  \right)\;.
\end{eqnarray}

The above projectors are also orthogonal under the scalar product
given by conjugating one of the two diagrams and summing over
color indices of the (anti)quarks.
We may therefore use them as (unnormalized) basis vectors
\begin{eqnarray}
  \mathbf{V}^6_{qq\rightarrow qq} &\propto& \raisebox{-0.5 \height}{\includegraphics[scale=0.35]{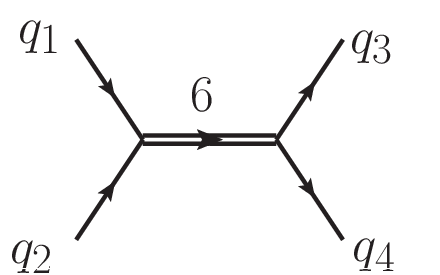}}
  \propto \mathbf{P}^6\nonumber \\
  \mathbf{V}^{\overline{3}}_{qq\rightarrow qq} &\propto&
  \raisebox{-0.5 \height}{\includegraphics[scale=0.35]{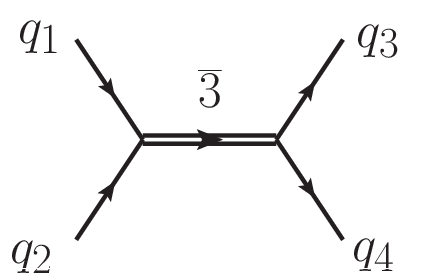}}
  \propto \mathbf{P}^{\overline{3}}\;.
\end{eqnarray}
An orthonormal basis can trivially be constructed, by using that
the basis vectors for a representation $\alpha$ square to
$\tr[\mathbf{P}^\alpha \mathbf{P}^\alpha ]=\tr[\mathbf{P}^\alpha]=d_\alpha$ whereas, for $\alpha\ne \beta$,
$\tr[\mathbf{P}^\alpha \mathbf{P}^\beta]=0$.

We note that there is no principal difference between an outgoing quark
and an incoming antiquark. The above basis can therefore also be used for
$\qqbar \rightarrow\qqbar$. Similarly, one can use a basis where a $\qqbar$-pair
is grouped to an octet or a singlet both for $\qqbar\rightarrow\qqbar$,
and $qq\rightarrow qq$.

For processes with gluons, the situation is more complicated, as it is
not straightforward to write down the projection operators
corresponding to the involved representations.
Say for example that we consider $g_1 g_2\rightarrow g_3 g_ 4$.
In principle we know the Clebsch-Gordan decomposition,
\begin{equation}
  \label{eq:8x8}
  8 \otimes 8=1 \oplus 8^a \oplus 8^s \oplus 10 \oplus \overline{10} \oplus 27\;,
\end{equation}
from Young tableau multiplication.
It is thereby known what projection operators to expect.
However, as opposed to the case of $qq \rightarrow qq$, this does not
reveal how to write down the corresponding
projection operators. For four gluons, the construction of projection operators
was first worked out for SU(3) in Ref. \cite{MacFarlane:1968vc}, then for
SU(N) \cite{Butera:1979na}, and
-- as it seems -- independently 
in Ref. \cite{Dokshitzer:2005ig} (see also \cite{Cvitanovic:1976am,Cvi08,Keppeler:2012ih}).
The constructions exploit symmetry and subtracts out parts proportional to
lower representations.

It should be remarked that the basis vectors are in general not
projection operators, but the projection operators for $n\rightarrow n$ gluons
can be used as a subset of the basis vectors for $2n$ gluons.
Instead one generally has to consider all admissible sets of representations
and vertices. This is applicable already to the four-gluon
case. A complete set of basis vectors spanning the overall (incoming
plus outgoing) color singlet space, involves not only the projection operators
corresponding to \eqref{eq:8x8}, but in general tensors mapping the representation
$\alpha$ to the representation $\alpha$, i.e. here 
also (orthogonal) vectors proportional
to $i f^{g_1 g_2 d} d^{d g_3 g_4}$ and $d^{g_1 g_2 d} if^{d g_3 g_4}$.
However,
charge conjugation symmetry 
can be used to put constraints on the basis vectors for the case of
only gluons \cite{Du:2015apa}. This implies that after all, for pure QCD, we do not need
to consider the basis vectors of form $i f^{g_1 g_2 d} d^{d g_3 g_4}$ and
$d^{g_1 g_2 d} if^{d g_3 g_4}$, as they cannot arise from the QCD Lagrangian.

The five-gluon
case was addressed in Ref. \cite{Sjodahl:2008fz}, and the case of a general number of gluons
was worked out later \cite{Keppeler:2012ih}, and results in basis vectors of form
$$
\raisebox{-0.5 \height}{\includegraphics[scale=0.4]{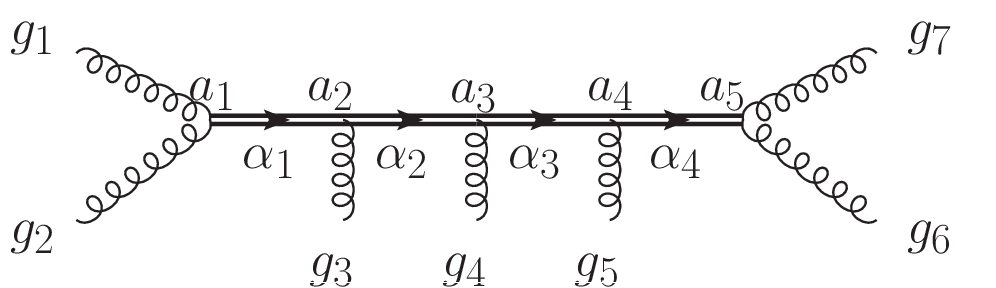}},
$$
where the double line $\alpha_n$ represent the overall representation
of the $n+1$ first gluons, and where we have taken care to write out
the vertex labels ($a_i$), since for a general representation $\alpha$, there
can be up to $\Nc-1$ instances of the representation $\alpha$ in
$A \otimes \alpha$, whereas all representations $\beta\ne \alpha$
come with multiplicity at most one, cf. \cite{Keppeler:2012ih}.

Using the birdtrack method (see Ref. \cite{Cvi08} or \cite{Keppeler:2017kwt,Peigne:2023iwm}), the scalar product of two
basis vectors, characterized by representations $\alpha_1,...,\alpha_n$
and  $\beta_1,...,\beta_n$ (and in general vertex numbers $a_1,...,a_n$ and $b_1,...,b_n$)
is given by 
\begin{equation}
  \raisebox{-0.5 \height}{\includegraphics[scale=0.4]{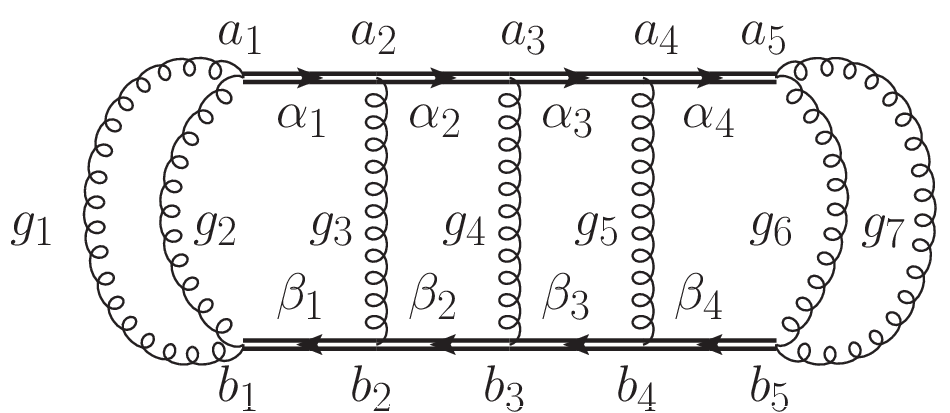}}\;,
\end{equation}
where the gluon dummy indices are implicitly summed over and could have been left out.
The orthogonality follows from
$$
\raisebox{-0.5 \height}{\includegraphics[scale=0.4]{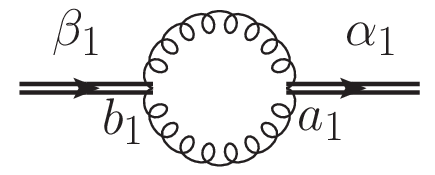}}
=
\frac{\raisebox{-0.7 \height}{\includegraphics[scale=0.4]{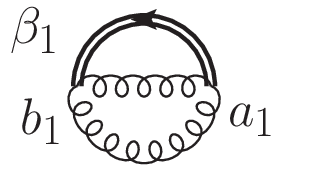}}} 
    {d_{\alpha_1}}
     {\raisebox{-0.2 \height}{\includegraphics[scale=0.4]{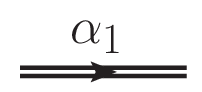}}}
     \delta_{{\alpha_1} {\beta_1}}\delta_{{a_1} {b_1}}\;,
$$
etc., where the first Kronecker delta signifies that the representations
$\alpha_1$ and $\beta_1$ have to be the same, and the second
states that the vertices have to be the same, for example
the symmetric and antisymmetric octets would be orthogonal
by construction.\footnote{We remark that the orthogonality of
the vertices is a choice; requiring the above basis vectors
to be orthogonal, we must choose the vertices (above $a_1$ and $b_1$) to be orthogonal.}
The above relation gives the orthogonality for $\alpha_1 \ne \beta_1$
(and for  $\alpha_1 = \beta_1$ but $a_1\ne b_1$). The orthogonality for $\alpha_2\ne\beta_2$
etc. follows by the same argument.

These gluon bases can also be used for constructing basis vectors
where quark-antiquark pairs are combined into singlets
or octets, for example
$$
\raisebox{-0.5 \height}{\includegraphics[scale=0.4]{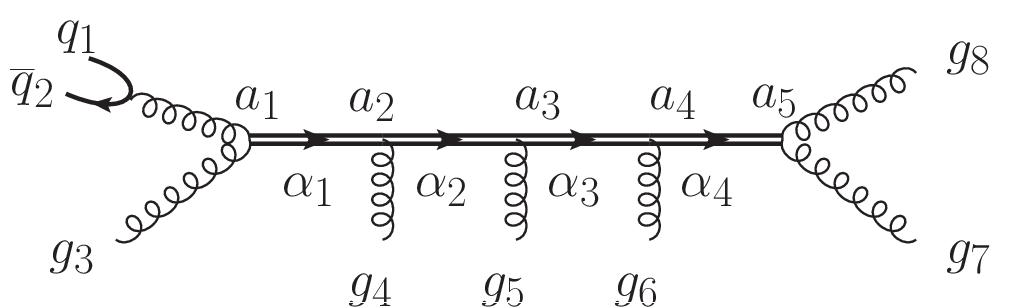}},
$$
for a single \qqbar-pair and six gluons, complemented with basis vectors where
the $\qqbar$-pair forms a singlet.

A more general type of multiplet basis, where quarks, antiquarks and gluons are
combined in arbitrary order onto a ``backbone'' chain of representations,
was outlined in Ref. \cite{Sjodahl:2018cca}, giving basis vectors of form
\begin{equation}
  \label{eq:MSJT}
  \raisebox{-0.5 \height}{\includegraphics[scale=0.4]{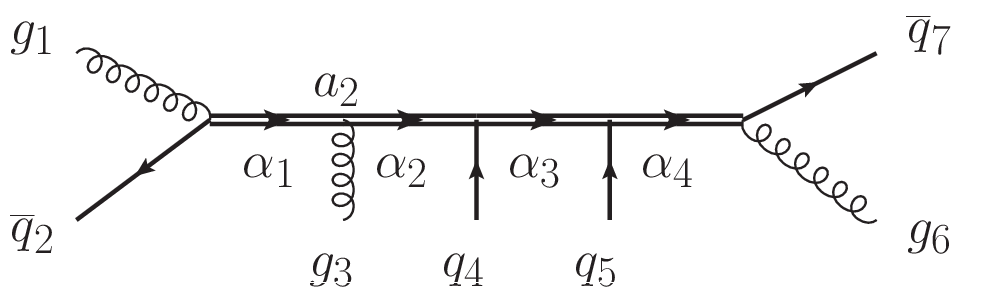}}\;,
\end{equation}
where the vertices with quarks only come with multiplicity one, and
therefore do not need any additional labels.

The orthogonality of these basis vectors follows by the
same argument as for the gluon bases above; unless all representations (and vertices)
are the same, the contraction will sooner or later result in a
vanishing Kronecker delta.

A characterizing feature of the basis vectors in  \eqref{eq:MSJT}
is that the partons (representations) can be attached to a ``backbone''
chain of representations in arbitrary (but fixed) order.\footnote{It should be
clear that the multiplet basis vectors are distinguished by the set of representations
(and vertices), as opposed to the trace bases where the different gluon orderings
result in different vectors.}
One can also
imagine multiplet bases, where partons are grouped together
in other ways, for example pairwise into representations which
then connect to each other.

\section{Orthogonal multiplet bases involving exotic representations}

We are now ready to construct orthogonal bases with exotic representations
by using the bases in \eqref{eq:MSJT}. The construction is straightforward:
keeping most of the representations untouched, we can group
quarks (or generally, quarks, antiquarks and gluons) into the
desired representation. For example, a basis vector with a single
sextet can be constructed using
\begin{eqnarray}
  \raisebox{-0.5 \height}{\includegraphics[scale=0.4]{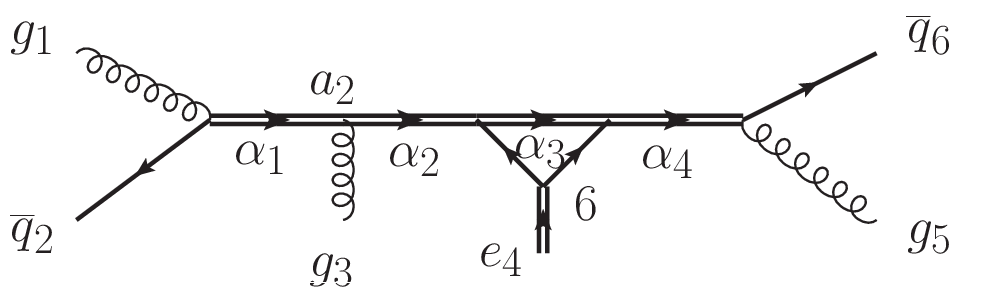}}
   =
   \frac{\raisebox{-0.5 \height}{\includegraphics[scale=0.45]{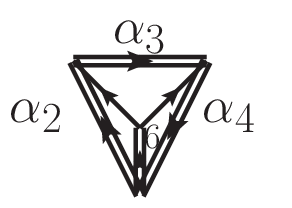}}}
       {\raisebox{-0.5 \height}{\includegraphics[scale=0.4]{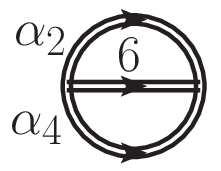}}}
   \raisebox{-0.5 \height}{\includegraphics[scale=0.4]{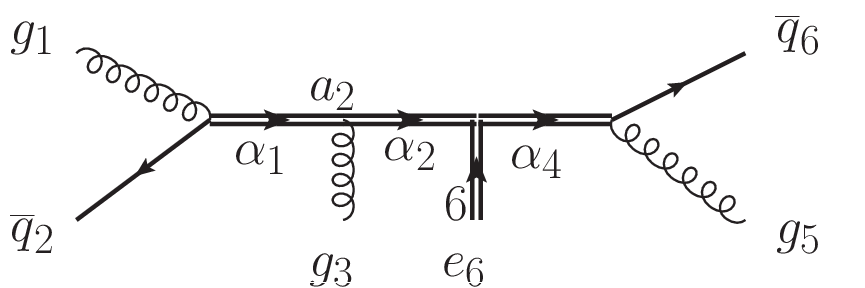}}\hspace*{-0.5 cm},
   \nonumber
\end{eqnarray}
where the sextet representation is enforced using the $\mathbf{P}^6$ symmetrizer
from \eqref{eq:projectors}, and all other vertices also can be
expanded out in terms of symmetrizers and antisymmetrizers,
paralleling the work in \cite{Keppeler:2012ih,Sjodahl:2018cca}.
If the process has an outgoing sextet, the procedure is repeated for a
neighboring pair of outgoing quarks.

Above, the constants in the numerator and denominator are Wigner $6j$ and $3j$
symbols (coefficients) respectively. As any fully contracted color structure,
they are just numbers.
The $3j$ symbol can --- if one prefers --- be normalized to one, by absorbing
a factor into the vertex definition. For the present construction,
we do, however, not need to know their value, the left-hand side of
the equation can be used as it is, or normalized for an orthonormal basis.

For the sextet representation, we have
$\alpha \otimes 6=\sum_{_i} \beta_i$, where each resulting representation $\beta_i$
only has multiplicity one.  We can therefore leave out the vertex label
also for this vertex.

For a theory with a decuplet, we can symmetrize in three quark
indices in a similar way, and again there will be only one instance of each
representation $\beta_i$ in $\alpha \otimes 10=\sum_{i} \beta_i$.
For more general representations, such as
$15=\begin{ytableau}  \ & \  &\\  \ \end{ytableau}$,
it can happen that
there is more than one instance of a resulting representation $\beta_i$,
and in this case one has to make sure to construct a complete set
of orthogonal vertices. For this, various intermediate representations
$\alpha_3$ can be used, and the vertices can be made orthogonal by standard
Gram-Schmidt orthogonalization.

Thus, utilizing the orthogonal multiplet bases in \eqref{eq:MSJT}
\cite{Sjodahl:2018cca},
it is comparatively straightforward to construct orthogonal
multiplet bases involving exotic representations.

\section{Conclusion and outlook}
\label{sec:conclusion}

In this letter we have shown how orthogonal multiplet basis vectors
involving exotic representations can be constructed,
essentially building on the
multiplet basis vectors in Ref. \cite{Sjodahl:2018cca}, and employing symmetrizers and
antisymmetrizers to enforce the exotic representations.

We remark that while this construction is in principle straightforward,
it will suffer from the same factorial scaling as
other multiplet bases constructed using symmetrizers and antisymmetrizers.
This is only a problem at construction time, but it still needs to be
addressed once. A possible algebraic route towards as solution
of the bad scaling was recently presented for the standard representations,
in Ref. \cite{chargeishvili2024automaticgenerationorthogonalmultiplet}.
It appears likely that a similar method can be employed
for the exotic representations addressed here.

An alternative route would be to use Wigner 6$j$
coefficients. For the standard model representations, and for
basis vectors of the form \eqref{eq:MSJT}, it was recently shown how
to calculate a sufficient set of 6$j$ coefficients,
directly in terms of other 6$j$ coefficients and dimensions of
representations \cite{Alcock-Zeilinger:2022hrk,Keppeler:2023msu}.
With such a set of $6j$ symbols at hand, it is not necessary to explicitly construct
basis vectors using symmetrizers and antisymmetrizers.
If such a route can be pursued also in this case, it is likely
more favorable.

\section*{Acknowledgments}

Stefan Keppeler, Thorsten Ohl and Simon Plätzer are thanked for comments on
the manuscript.
MS acknowledges support by the Swedish Research Council (contract
number 2016-05996), as well as by the European Union’s Horizon 2020
research and innovation programme (grant agreement No 668679).

\bibliographystyle{elsarticle-num}

\bibliography{refs}

\end{document}